\documentclass[preprint, 12pt]{elsarticle}
\pdfoutput=1

\usepackage{amssymb}
\usepackage{amsthm}

\usepackage{amsmath}
\usepackage{amssymb}
\usepackage{amsfonts}
\usepackage{epsfig}
\usepackage{graphics}
\usepackage{color}

\newcommand\XOR{\mathbin{\char`\^}}

\journal{Computer Physics Communications}

\begin{document}

\begin{frontmatter}



\title{Symbolic integration of a product of two spherical bessel functions with an additional
exponential and polynomial factor}

\author[nscl,msu]{B.~Gebremariam\corref{cor1}}
\ead{gebremar@nscl.msu.edu}
\author[nscl,msu,sac]{T.~Duguet\corref{cor2}}
\ead{thomas.duguet@cea.fr}
\author[nscl,msu]{S.~K.~Bogner\corref{cor3}}
\ead{bogner@nscl.msu.edu}
\address[nscl]{National Superconducting
Cyclotron Laboratory, 1 Cyclotron Laboratory, East-Lansing, MI
48824, USA}
\address[msu]{Department of Physics and Astronomy,
Michigan State University, East Lansing, MI 48824, USA}
\address[sac]{CEA, Centre de Saclay, IRFU/Service de Physique
Nucl{\'e}aire, F-91191 Gif-sur-Yvette, France}

\begin{abstract}
We present a mathematica package that performs the symbolic calculation of integrals of the form
\begin{equation}
\int^{\infty}_0 e^{-x/u} x^n j_{\nu} (x) j_{\mu} (x) dx \,
\end{equation}
where $j_{\nu} (x)$ and $j_{\mu} (x)$  denote spherical Bessel functions of integer orders, with $\nu \ge 0$ and $\mu \ge 0$.
With the real parameter $u>0$ and the integer $n$, convergence of the integral requires that $n+\nu +\mu  \ge 0$. The package provides analytical result for the integral in its most simplified form. The novel symbolic method employed enables
the calculation of a large number of integrals of the above form in a fraction of the time required for conventional numerical and Mathematica based brute-force methods. We test the accuracy of such analytical expressions by comparing the results with their numerical counterparts.
\end{abstract}

\end{frontmatter}



{\em PROGRAM SUMMARY}

\begin{small}
\noindent
{\em Program Title:} SymbBesselJInteg    \\
{\em Catalogue identifier:}                \\
{\em Licensing provisions:}                \\
{\em Programming language:}  Mathematica 7.1\\
{\em Computer:} Any computer running Mathematica 6.0 and later versions\\
{\em Operating system: Windows Xp, Linux/Unix} \\
{\em RAM:} 256 Mb \\
{\em Classification:} 5\\
{\em Nature of problem:} Integration, both analytical and numerical, of products of two spherical bessel functions with an exponential and polynomial multiplying factor can be a
very complex task depending on the orders of the spherical bessel functions. The Mathematica package discussed in this paper solves this problem
using a novel symbolic approach.\\
{\em Solution method:} The problem is first cast into a related limit problem which can be broken into two related subproblems involving
exponential and exponential integral functions. Solving the cores of each subproblem symbolically sets the stage for an involved expression tree parsing and manipulation to obtain the most simplified analytic expression for the initial problem.\\
{\em Running time:} 1 min for typical values of the arguments and can be several mins for large values of the input variables. For the test data included, about an hour.
\end{small}




\section{Introduction}\label{Intro}
Integrals of products of spherical bessel functions multiplied by an exponential times a polynomial occur in a variety
of fields. Nuclear physics, scattering theory, calculation of Feynman diagrams, hydrodynamics and elasticity theory are notable examples where such integrals appear~\cite{Davis, Groote, Stone, Tanzosh,Tezer, Whelan, Conway,Mobilia}.
Several authors have proposed various numerical modules that can be used to obtain accurate numerical integrations
of these functions ~\cite{Deun, Singh}. While most of these deal with general Bessel functions, spherical bessel functions
actually appear in most physical problems that have spherical symmetry. A case in point is scattering problems.

Recently, the application of the density matrix expansion ~\cite{Negele, Gebremariam1} to the HF energy
obtained from chiral EFT two- and three-nucleon interactions at N$^2$LO ~\cite{Bogner, Gebremariam2, Gebremariam3, Gebremariam4} required analytical expressions for integrals of the form
\begin{equation}\label{Idef}
I(u, n,\mu, \nu) \,= \,\int^{\infty}_0 e^{-x/u} x^n j_{\nu} (x) j_{\mu} (x) dx \,
\end{equation}
where $j_{\nu} (x)$ and $j_{\mu} (x)$  denote spherical Bessel functions of integer orders, with $\nu \ge 0$ and $\mu \ge 0$.
The real parameter $u$ is such that $u>0$ and $n$ is an integer constant. Using the asymptotic properties of spherical bessel functions and convergence requirement as the argument $x$ goes to zero, one obtains the constraint $n+\nu +\mu \ge 0$.

 In addition to needing analytical expressions of $I(u, n,\mu, \nu)$, the application of Ref.~\cite{Gebremariam2} required
the evaluation of several hundred of such integrals. Furthermore, in a large number of cases, the  direct implementation of these
integrals in Mathematica fails to give closed expressions. Finally, it should be mentioned that as the order of the spherical bessel functions
$\nu$ and $\mu$ as well as the exponent $n$ grow, the expressions one obtains for the integrals from the direct implementation become unwieldy
if and when one has a closed expression for those integrals. All these problems necessitated the development of a dedicated Mathematica package that is the subject of the present paper.
The paper is structured as follows: in the next section we discuss the main symbolic algorithm and its limitations. This is followed by section ~\ref{Mathe-Implem} that presents its Mathematica implementation. Section ~\ref{moreonpack} is devoted to
actual details of the package and to an illustrative example that shows how one can use the developed package. The last two sections contain the
conclusions and the appendix.

\section{Symbolic Integration}\label{SymbIntAlgo}
In this paper, the term symbolic integration signifies a method of computing complicated integrals using the
techniques of expression tree parsing, extraction, deletion and replacement with the aim of obtaining a simplified expression. The final expression must be the same as the one obtained analytically, if possible, whereas for given input values, it should evaluate to the same value as the corresponding numerical integration.

\subsection{Recasting the problem}
The starting point for the symbolic integration algorithm is the definition
\begin{equation}\label{limitdef}
F(y, u, n,\mu, \nu) \, \equiv \,  \,\int^{y}_0 e^{-x/u} x^n j_{\nu} (x) j_{\mu} (x) dx\,,
\end{equation}
that allows the rewriting of Eq.\eqref{Idef} as
\begin{equation}
I(u, n, \nu, \mu) \, = \, \lim_{y\to\infty} F(y, u , n, \mu,\nu)\,,
\end{equation}
which is valid for all convergent integrals. The essence of such a simple step relates to the fact that
Mathematica can evaluate $F(y, u, n,\mu, \nu)$ for all cases of interest (note the specified domain).
One can separate the resulting expression for $F(y, u, n,\mu, \nu)$ into a part that contains an explicit $y$ dependence and
another part without $y$ dependence, viz,
\begin{equation}
F(y, u, n,\mu, \nu) \, \equiv  \, G(y, u, n,\mu, \nu)  + H (u, n,\mu, \nu) \,.
\end{equation}

Since the most troublesome part is the evaluation of the limit as $y$ goes to infinity, the next several steps
pertain to the symbolic evaluation of this limit. Direct
evaluation of the limit of $G(y,u,n,\mu,\nu) $ as $y$ goes to infinitydoes not seem to work or takes an impractical
amount of time for a large part of the
parameter space $\,(n,\,\mu,\,\nu)$. This is what necessitated the development of the core of the symbolic algorithm.

\subsection{The core of the algorithm}
Extensive tests performed for a large number of combinations of parameters $n, \mu, \nu$ show that Mathematica always expresses $G(y, u, n,\mu, \nu)$ in terms of the exponential integral function and exponential functions of the form $e^{f(y,u)}$ where
$f(y, u)$ is some function. The exponential integral function is defined formally as ~\cite{Abramowitz, Wolfram}\footnote{There is a sign difference between the definitions given in Ref.~\cite{Abramowitz} and Ref.~\cite{Wolfram}. That of Ref.~\cite{Wolfram} is adopted in this work.}
\begin{equation}\label{expinteg}
Ei(x) \, = \, -\int^{\infty}_{-x} \frac{e^{-t}}{t} d t\,.
\end{equation}
Furthermore, one finds that the y-dependence of $G(y, u, n,\mu, \nu)$ is completely absorbed by $f(y,u)$ and by the exponential integral functions. Additionally, $G(y, u, n,\mu, \nu)$ does not contain products of exponential functions and exponential integral functions. Hence, the generic expression that Mathematica produces for $G(y, u, n,\mu, \nu)$ can be written as
\begin{equation}\label{Gdef}
G(y, u, n,\mu, \nu)  \, =\, \sum^{N}_{i=1} a_i(u) e^{f_i(y,u)} + \sum^{M}_{j=1} b_j(u) Ei(g_j(y,u)) \,,
\end{equation}
where $N$ and $M$ are integers whereas $a_i(u),\, f_i(y,u),\,b_j(u),\, g_j(y,u)$ denote some unspecified auxiliary functions. At this point, the algorithm does not require the exact form of these auxiliary functions to be specified.
Finally, we note that the structure of $G(y, u, n,\mu, \nu)$ as given by Eq.\eqref{Gdef} can be understood adequately by applying the method of partial integration to Eq.\eqref{limitdef}.

In the evaluation of the limit of $G(y, u, n,\mu, \nu)$ as y goes to infinity, the first group of terms terms in Eq.\eqref{Gdef}, viz, the ones that contain exponential functions can be simplified directly. Thus the problem narrows down to the evaluation of the limit of the second set of terms in Eq.\eqref{Gdef}, specifically the ones containing exponential integral functions. In line with that, the following observations are integral parts of the algorithm:
\begin{enumerate}[(i)]
\item There appear only three distinct exponential integral functions, i.e. $M = 3$. Their argument functions, $g_j(y,u)$, have
the form
\begin{eqnarray}\label{constraint1}
g_1(y,u)&=& - \frac{y}{u}\,,\nonumber\\
g_2(y,u)&=&-i 2 y- \frac{y}{u}\,,\nonumber\\
g_3(y,u)&=& i 2 y - \frac{y}{u} \,.
\end{eqnarray}
\item The real parts of the $b_j(u)$ coefficient functions satisfy the constraints
\begin{eqnarray}\label{constraint2}
Re\biggl[\sum^{3}_{j=1} b_j(u)\biggr] &=& 0 \,,\nonumber\\
Re\bigl[b_2(u)\bigr] &=& Re\bigl[b_3(u)\bigr]\,,
\end{eqnarray}
\item The imaginary parts of the $b_j(u)$ coefficient functions satisfy
\begin{eqnarray}\label{constraint3}
Im\bigl[b_1(u)\bigr]&=&0\,,\nonumber\\
Im\bigl[ b_2(u)\bigr] &=& - Im\bigl[b_3(u)\bigr] \,,
\end{eqnarray}
\end{enumerate}
where $Re$ and $Im$ extract the real and imaginary parts of the arguments respectively. As discussed in the implementation section (see section ~\ref{Mathe-Implem}), all these conjectures can be verified during the program's progress. Hence, in the evaluation of the limit of $G(y, u, n,\mu, \nu)$ as y goes to infinity and after all the symbolic manipulations that need to be performed are marked out, one is left with the evaluation of the limits
\begin{eqnarray}
h_1 (u) &= &\lim_{y\to\infty}\biggl[ Ei\biggl(-i 2 y -\frac{y}{u}\biggr) +  Ei\biggl(i 2 y-\frac{y}{u}\biggr)
-2 \, Ei\biggl(-\frac{y}{u}\biggr)\biggr]\label{h1}\,,\\
h_2 (u)  &=& \lim_{y\to\infty}\biggl[ Ei\biggl(-i 2 y -\frac{y}{u} \biggr) -  Ei\biggl(i 2 y -\frac{y}{u} \biggr)\biggr]\label{h2} \,.
\end{eqnarray}
Analytic evaluation of these limits as outlined in the appendix (see section~\ref{appendix1}) yield the results
$h_1(u) =0$ and $h_2(u) = -i\, 2 \pi$. As mentioned in section \ref{Mathe-Implem}, Mathematica expectedly gives
the same answer though, in a very complicated form as demonstrated in Eqs. \eqref{h1unsimplified} and \eqref{h2unsimplified}. Refer to section \ref{simplifier} for the role of the simplifier part
of the Mathematica package in this context. The subsequent steps of the algorithm comprise making use of the listed conjectures
and a host of symbolic manipulation operations to obtain the most simplified expression for $I(u, n, \mu, \nu)$.
\subsection{Comments on the algorithm}
The symbolic integration algorithm given in the previous section is mostly based on Mathematica "symbolic experiments" using a large number of values for the input variables $n,\,\mu\,$ and $\,\nu$. As such, most part of the algorithm is based on
empirical observation. Even so, all of the conjectures in the algorithm are verifiable at each stage of the symbolic computation, as discussed in section \ref{Mathe-Implem}. It would of course be beneficial to verify those conjectures analytically as our intensive and
systematic testing on a vast number of values for the input variables did not locate a single exception to these conjectures.

\section{Mathematica Implementation}\label{Mathe-Implem}
The Mathematica implementation of the symbolic algorithm consists of three main parts: parser, simplifier and checker.
In the following, we discuss the role and the implementation of each part.
\subsection{Expression parser}\label{parser}
The first task of the expression parser is to take $F(y,u,n,\mu,\nu)$ as an input and
extract $G(y,u,n,\mu.\nu)$ and $H(y,u,n,\mu.\nu)$. Hence, we define the input $F(y,u,n,\mu.\nu)$ which we call {\em Fint} as
\begin{eqnarray}
\text{{\em Fint}}&  = &\text{\em Expand[Integrate[Exp[-x/u] x$\XOR$n  SphericalBesselJ[$\mu$, x ]} \nonumber\\
&&\, \text{\em SphericalBesselJ[$\nu$, x], Assumptions$\rightarrow$\{Im[u]==0, u $>$ 0\}]]} \nonumber,
\end{eqnarray}
where the expression is expanded to make the expression tree available for the subsequent parsing steps. The key Mathematica functions to extract $G(y,u,n,\mu.\nu)$ and $H(y,u,n,\mu.\nu)$ from {\em{Fint}} are  {\em{Position}} and {\em{Head}}, along with a few other helper functions.

In the next several steps, all the conjectures mentioned in section \ref{SymbIntAlgo} are verified by parsing and traversing the
expression tree. In connection with this, the Mathematica function {\em{Coefficient}} turns out to be handy. Specifically, the conjectures that
need verification are (i) the existence of only exponential and exponential integral functions in $G(y,u,n,\mu,\nu)$
(ii) that the terms containing the exponential function vanishe in the limit of y going to infinity (iii) the set of conditions specified in Eqs. \eqref{constraint1}, \eqref{constraint2} and \eqref{constraint3}. The expression parser aborts the calculation and produces error message if any one of the above conjectures fails to hold. As a side note, we mention that for the domain specified in section ~\ref{Intro}, all the conjectures have been verified (see section ~\ref{result-validity}). In the following step, the expression parser makes use of the two basic limits identified in Eqs.\eqref{h1} and \eqref{h2} to rewrite $I(u,n,\mu,\nu)$ as
\begin{equation}\label{final-parser-eqn}
 I(u,n,\mu,\nu) =  H(u,n,\mu,\nu) + C_1(u,n,\mu,\nu) h_1(u) + C_2(u,n,\mu,\nu) h_2(u)\,,
\end{equation}
where the coefficient functions $C_1(u,n,\mu,\nu)$ and $C_2(u,n,\mu,\nu)$ are determined using the {\em{Coefficient}} Mathematica functions and some additional parsing. The corresponding Mathematica code for Eq.\eqref{final-parser-eqn} reads
\begin{equation}
\text{\em{Iint = HIint +  Coef1*Lim1 + Coef2*Lim2   }}\nonumber\,,
\end{equation}
where {\em{Iint}} is the Mathematica variable for $I(u,n,\mu,\nu)$, {\em{HIint}} is the Mathematica variable for $H(u,n,\mu,\nu)$, {\em{Coef1}} is that of  $C_1(y,u,n,\mu,\nu)$,
{\em{Lim1}} is for $h_1(u)$, {\em{Coef2}} is that of  $C_2(y,u,n,\mu,\nu)$ and
{\em{Lim2}} is for $h_2(u)$.

Paving the way for the discussion of the simplifier part of the package, note that Mathematica produces $h_1(u)$
and $h_2(u)$, prior to any simplification, under the form

\begin{eqnarray}\label{h1unsimplified}
h_1(u) &=&\frac{1}{2 (1+4 u^2)^2}\biggl(\,\text{Log}[-2 i-1/u]+8 u^2 \,\text{Log}[-2 i-1/u]\nonumber\\
&& +16 u^4 \,\text{Log}[-2 i-1/u]+8 u^2 \,\text{Log}[2 i-1/u]+16 u^4 \,\text{Log}[2 i-1/u]\nonumber\\
&& -\,\text{Log}[-2 i+1/u]-8 u^2 \,\text{Log}[-2 i+1/u]-16 u^4 \,\text{Log}[-2 i+1/u]\nonumber\\
&&  +8 u^2 \,\text{Log}[\frac{i u}{i-2 u}]-8 u^2 \,\text{Log}[2 i+1/u]-16 u^4 \,\text{Log}[2 i+1/u]\nonumber\\
&& +8 u^2 \,\text{Log}[-1-2 i u]+16 u^4 \,\text{Log}[-1-2 i u]+8 u^2 \,\text{Log}[-1+2 i u]\nonumber\\
&& +16 u^4 \,\text{Log}[-1+2 i u]-16 u^2 \,\text{Log}[u]-32 u^4 \,\text{Log}[u]+\,\text{Log}[\frac{i u}{i-2 u}]\nonumber\\
&&-\,\text{Log}[2 i+1/u]+\,\text{Log}[\frac{i u}{i+2 u}]  -\,\text{Log}[\frac{i u}{-i+2 u}] -\,\text{Log}[-\frac{i u}{i+2 u}]\nonumber\\
&&+16 u^4 \,\text{Log}[\frac{i u}{i-2 u}] +8 u^2 \,\text{Log}[\frac{i u}{i+2 u}]+16 u^4 \,
\text{Log}[\frac{i u}{i+2 u}]\nonumber\\
&&+\,\text{Log}[2 i-1/u]\,\biggr)\,,
\end{eqnarray}
and
\begin{eqnarray}\label{h2unsimplified}
h_2 (u)&=&\frac{1}{2 (1+4 u^2)^2}\biggl(\,8 u^2 \,\text{log}[-2 i-1/u]+16 u^4 \,\text{log}[-2 i-1/u]\nonumber\\
&&-\,\text{log}[2 i-1/u]-8 u^2 \,\text{log}[2 i-1/u]-16 u^4 \,\text{log}[2 i-1/u]\nonumber\\
&&
+\,\text{log}[-2 i+1/u]+8 u^2 \,\text{log}[-2 i+1/u]+16 u^4 \,\text{log}[-2 i+1/u]\nonumber\\
&&-\,\text{log}[2 i+1/u]-8 u^2 \,\text{log}[2 i+1/u]-16 u^4 \,\text{log}[2 i+1/u]\nonumber\\
&& +8 u^2 \,\text{log}[-1-2 i u]+16 u^4 \,\text{log}[-1-2 i u]-8 u^2 \,\text{log}[-1+2 i u]\nonumber\\
&&-16 u^4 \,\text{log}[-1+2 i u]+\,\text{log}[\frac{i u}{i-2 u}]+8 u^2 \,\text{log}[\frac{i u}{i-2 u}]\nonumber\\
&&+16 u^4 \,\text{log}[\frac{i u}{i-2 u}]-\,\text{log}[\frac{i u}{-i+2 u}]+\,\text{log}[-\frac{i u}{i+2 u}]\nonumber\\
&&-\,\text{log}[\frac{i u}{i+2 u}]-8 u^2 \,\text{log}[\frac{i u}{i+2 u}]-16 u^4 \,\text{log}[\frac{i u}{i+2 u}]\nonumber\\
&& + \text{log}[-2 i-1/u]\biggr)\,,
\end{eqnarray}
which are far more complex than the simple expressions that obtained from the ouputs of the simplifier or analytic derivation
as discussed in appendix ~\ref{appendix1}.

To summarize, the key task of the the expression parser is to implement the Mathematica code outlined just after Eq.\eqref{final-parser-eqn}, thereby arriving at a completely analytical but unnecessarily complicated expression for $I(u, n,\mu,\nu)$. Details of the implementation can be found in the submitted Mathematica code.

\subsection{Expression simplifier}\label{simplifier}
At this point, we have at hand analytic expressions for $I(u, n,\mu,\nu)$. However, as can be expected from the previous discussion, the expressions are extremely complicated and barely useful. To illustrate, the expression one obtains for $I(u,-1,1,1)$ is

{\allowdisplaybreaks[4]
\begin{align}\label{SampleI}
I(u,-1,1,1 ) &=\frac{1}{4}-\frac{1}{24 u^2}+\frac{\,\text{Log}[-2 i-\frac{1}{u}]}{192 u^4}+\frac{\,\text{Log}[-2 i-\frac{1}{u}]}{16 u^2}+\frac{i \,\text{Log}[-2 i-\frac{1}{u}]}{12 u}\,\nonumber \\
&+\frac{\,\text{Log}[2 i-\frac{1}{u}]}{192 u^4} +\frac{\,\text{Log}[2 i-\frac{1}{u}]}{16 u^2}-\frac{i \,\text{Log}[2 i-\frac{1}{u}]}{12 u}-\frac{\,\text{Log}[-\frac{1}{u}]}{96 u^4}\,\nonumber \\
&-\frac{\,\text{Log}[-\frac{1}{u}]}{8 u^2}+\frac{\,\text{Log}[-u]}{96 u^4}+\frac{\,\text{Log}[-u]}{8 u^2} -\frac{\,\text{Log}[\frac{u}{-1-2 i u}]}{192 u^4}
\,\nonumber \\
&-\frac{\,\text{Log}[\frac{u}{-1-2 i u}]}{16 u^2}-\frac{i \,\text{Log}[\frac{u}{-1-2 i u}]}{12 u}-\frac{\,\text{Log}[\frac{u}{-1+2 i u}]}{192 u^4}
-\frac{\,\text{Log}[\frac{u}{-1+2 i u}]}{16 u^2} \,\nonumber \\
&+\frac{i \,\text{Log}[\frac{u}{-1+2 i u}]}{12 u}+\,\text{Log}[\frac{i u}{i-2 u}] -8 u^2 \,\text{Log}[\frac{i u}{i+2 u}]
\,\nonumber \\
&+8 u^2 \,\text{Log}[-2 i-\frac{1}{u}]\,\nonumber \\
&-\frac{i}{12 u (1+4 u^2)^2} \biggl(\,\text{Log}[-2 i-\frac{1}{u}]+8 u^2 \,\text{Log}[-2 i-\frac{1}{u}] \,\nonumber \\
&+16 u^4 \,\text{Log}[-2 i-\frac{1}{u}]-\,\text{Log}[2 i-\frac{1}{u}]-8 u^2 \,\text{Log}[2 i-\frac{1}{u}]\nonumber \\
&-16 u^4 \,\text{Log}[2 i-\frac{1}{u}] \,+\,\text{Log}[-2 i+\frac{1}{u}]+8 u^2 \,\text{Log}[-2 i+\frac{1}{u}]\nonumber \\
&+16 u^4 \,\text{Log}[-2 i+\frac{1}{u}]-\,\text{Log}[2 i+\frac{1}{u}]
-8 u^2 \,\text{Log}[2 i+\frac{1}{u}]\nonumber \\
&-16 u^4 \,\text{Log}[2 i+\frac{1}{u}]+8 u^2 \,\text{Log}[-1-2 i u]+16 u^4 \,\text{Log}[-1-2 i u] \,\nonumber \\
&-8 u^2 \,\text{Log}[-1+2 i u]-16 u^4 \,\text{Log}[-1+2 i u]+\,\text{Log}[\frac{i u}{i-2 u}]\nonumber \\
&+8 u^2 \,\text{Log}[\frac{i u}{i-2 u}] +16 u^4 \,\text{Log}[\frac{i u}{i-2 u}]-\,\text{Log}[\frac{i u}{-i+2 u}]\nonumber \\
&+\,\text{Log}[-\frac{i u}{i+2 u}]-\,\text{Log}[\frac{i u}{i+2 u}]-16 u^4 \,\text{Log}[\frac{i u}{i+2 u}]\biggr)\,\nonumber \\
&+\frac{1}{96 u^4 (1+4 u^2)^2}(1+12 u^2) \biggl(\,\text{Log}[-2 i-\frac{1}{u}]+16 u^4 \,\text{Log}[-2 i-\frac{1}{u}]\,\nonumber \\
&+\,\text{Log}[2 i-\frac{1}{u}]+8 u^2 \,\text{Log}[2 i-\frac{1}{u}]+16 u^4 \,\text{Log}[2 i-\frac{1}{u}] \,\nonumber \\
&-\,\text{Log}[-2 i+\frac{1}{u}]-8 u^2 \,\text{Log}[-2 i+\frac{1}{u}]-16 u^4 \,\text{Log}[-2 i+\frac{1}{u}]\,\nonumber \\
&-\,\text{Log}[2 i+\frac{1}{u}] -8 u^2 \,\text{Log}[2 i+\frac{1}{u}]-16 u^4 \,\text{Log}[2 i+\frac{1}{u}]\,\nonumber \\
&+8 u^2 \,\text{Log}[-1-2 i u]+16 u^4 \,\text{Log}[-1-2 i u] +8 u^2 \,\text{Log}[-1+2 i u]\,\nonumber \\
&+16 u^4 \,\text{Log}[-1+2 i u]-16 u^2 \,\text{Log}[u]-32 u^4 \,\text{Log}[u] \,\nonumber \\
&+8 u^2 \,\text{Log}[\frac{i u}{i-2 u}]+16 u^4 \,\text{Log}[\frac{i u}{i-2 u}]-\,\text{Log}[\frac{i u}{-i+2 u}]\,\nonumber \\
&-\,\text{Log}[-\frac{i u}{i+2 u}] +\,\text{Log}[\frac{i u}{i+2 u}]+8 u^2 \,\text{Log}[\frac{i u}{i+2 u}]\,\nonumber \\
&+16 u^4 \,\text{Log}[\frac{i u}{i+2 u}]\biggr)\,.
\end{align}
}

It should be noted that, the above expression is obtained for relatively small values of the arguments $n,\,\mu$  and $\nu$.
As these arguments increase, the resulting expression becomes much more unwieldy. The expression simplifier makes use of the power of Mathematica's symbolic manipulation functions to obtain the
most simplified expression for $I(u,n,\mu,\nu)$. This task is complicated by the branch-cuts of the complex logarithm function.
Due to this and the complexity of the starting expression, direct simplification does not seem to yield the most simplified expression. Rather we use expression tree parsing and manipulation that
consists of the following steps: (i) simplify each $Log$ expression in terms of the three elementary $Logs$ that constitute the entire $Log$ dependence: $\text{Log}[i 2 u + 1], \,
\text{Log}[i 2 u - 1]$  and $\text{Log}[u]$. This step is not as trivial as it seems due to the need to handle the complex arguments properly. (ii) Check the coefficients of the $\text{Log}[i 2 u + 1]$ and $\text{Log}[i 2 u - 1]$
are the same and real (as $I(u,n,\mu,\nu)$ is known to be real) and combine them into $\text{Log}[4 u^2 + 1] + \text{Log}[-1]$, and
(iii) finally check that the whole expression is real. The Mathematica functions that are most important for such parsing and simplification are the {\em{Numerator}}, {\em{Denominator}}, {\em{Extract}}, {\em{Together}} and {\em{Head}}. For the actual implementation, refer to the submitted Mathematica code. To demonstrate the power of this technique, taking $I(u,-1,1,1)$ as given by
Eq.\eqref{SampleI} as input, the expression simplifier produces the following expression
\begin{eqnarray}\label{SampleIAnalytical}
I(u,-1,1,1)\,&=&\,\frac{1}{96 u^4} \,\biggl(4 u^2 (-1+6 u^2)\,-\,32 u^3 \,\text{ArcTan}[2 u]\nonumber\\
&& \,+\,(1+12 u^2) \,\text{Log}[1+4 u^2] \biggr)\,.
\end{eqnarray}
As a final test, we note that the expression simplifier indeed recovers the simplified results for $h_1(u)$ and $h_2(u)$
(starting from Eqs.\eqref{h1unsimplified}  and \eqref{h2unsimplified}) which are also
one obtained analytically in appendix \ref{appendix1}. In the next section we discuss
the final part of the submitted Mathematica code.

\subsection{Expression checker}\label{checker}
From the implementation point of view, this is the simplest part of the code.
The expression checker tests the accuracy of the simplified analytical expressions for $I(u, n,\mu,\nu)$
against a high precision numerical calculation of the same quantity. Specifically, we make used of the {\em{NIntegrate}} function in Mathematica
with a precision of $10$ decimal digits and compare the output of the symbolic with the numerical computation
for a range of values of $u$. Hence, for the sample analytical expression given in Eq. \eqref{SampleIAnalytical}, the numerical
counterpart is produced by
\begin{eqnarray}
\text{\em templist}& =& \text{\em NIntegrate[Exp[-x/tempu] x$\XOR$(-1)  SphericalBesselJ[1, x ]} \nonumber \\
&&\text{\em SphericalBesselJ[1, x], \{x,0,1000*tempu\}, PrecisionGoal$\rightarrow$10] }\,,\nonumber
\end{eqnarray}
where {\em templist} is the value of the numerical integration for some $u=\text{\em tempu}$. We discuss the result in
section ~\ref{result-validity}.

\section{More on the Mathematica package}\label{moreonpack}
There are two files submitted to the electronic library. The first file is the actual Mathematica package {\emph{SymbBesselJInteg.m}}
that contains the implementation of symbolic integration algorithm and related accuracy checking function. In the second file, we have
implemented an extensive test of the algorithm.
\subsection{Program structure}
There are three main functions made accessible in the package. These are
\begin{enumerate}[(i)]
\item {\em ParseCalcIInt[$u$, $n$, $\mu$,  $\nu$]} which takes four arguments that correspond to the four variables in
Eq. \eqref{Idef}. The function calculates and returns $I(u,n,\mu,\nu)$ without simplification, in essence implementing the
expression parser discussed in section \ref{parser}.

\item {\em Simplifier[$u$, $Iint$]} whose second argument should be the exact expression returned by the {\emph{ParseCalcInt}}
 function and while the first one is the basic variable of the problem $u$. The function essentially implements the expression simplifier
 discussed in section \ref{simplifier}. Its return value is the most simplified expression corresponding to the input expression.

 \item {\em CheckAnalyNumer[$u$, $n$, $\mu$, $\nu$, $uList$, $Iint$]} is the function that tests the accuracy of the symbolic
 integration algorithm against a numerical result of the same quantity. It takes six arguments. While the meaning of the first
 four arguments should be self-explanatory, the fifth one, $uList$, refers to a list containing $u$ values for which the test
 is going to be carried out. The last argument, $Iint$, refers to either the output of the {\em ParseCalcInt} function
 or that of {\em Simplifier}, depending on which code segment is being tested. In short, this function implements the expression checker
 discussed in section \ref{checker}. Finally, the function returns a list containing both the real and imaginary\footnote{The imaginary part is expected to be zero.} parts of both the analytical expression and the corresponding numerical result.
\end{enumerate}

In the following, we illustrate how one can make use of the three functions in the package to calculate Eq.\eqref{Idef}
and to verify the accuracy.
\subsection{Illustrative example}
In order to demonstrate how to use the package, we calculate $I(u,-3,3,3)$. First, one needs to load the package into the
Mathematica session. Assuming the package is in the "temp" directory, this is accomplished by
\begin{align}
\text{In[1]:}&=\text{\em SetDirectory["/temp"];} \nonumber\\
\text{In[2]:}&=\text{\em Needs["SymbBesselJInteg`"];}\nonumber
\end{align}
To get a list and description of the functions made available in the package, one can use the command
\begin{align}
\text{In[3]:}&=\text{\em ?SymbBesselJInteg`*} \nonumber
\end{align}
In the next step, one should clear the basic symbol which we call $u$ (note that any variable name is valid except that one has
to use it consistently) and call the {\em ParseCalcInt} function as
\begin{align}
\text{In[4]:}&=\text{\em Clear[u, n, $\mu$ , $\nu$ , uList];}\nonumber\\
\text{In[5]:}&=\text{\em n = -3; $\mu$ = 3; $\nu$ = 3;}\nonumber\\
\text{In[6]:}&=\text{\em Iint = ParseCalcInt [u ,n, $\mu$, $\nu$];}\nonumber
\end{align}
At this point, the variable {\em Iint} has the complicated analytical expression for  $I(u,-3,3,3)$. The simplification
proceeds by invoking
\begin{center}
\text{In[7]:}=\text{\em IintSimplified = Simplifier[u ,Iint]}\\
\end{center}
Thus {\em IintSimplified} contains the simplified expression. For the case at hand, viz, $I(u,-3,3,3)$, it reads
\begin{eqnarray}
\text{Out[7]:}&=& \frac{1}{967680 u^{10}}\biggl[4 u^2\, \bigl(-15-240 u^2-1556 u^4+4272 u^6+672 u^8\bigr) \nonumber\\
&& -1536 u^7 \,\bigl(15+4 u^2\bigr)\, \text{ArcTan}[2 u]+3 \bigl(5+90
u^2+672 u^4+3360 u^6\bigr)\,\nonumber\\
&&\times  \text{Log}\bigl[1+4 u^2\bigr]\biggr]\,.
\end{eqnarray}
Finally, to check the accuracy, one starts by definining a list of values of $u$ as, for example,
\begin{eqnarray}
\text{In[8]:}&=\text{\em uList = \{0.5, 1.0, 2.0, 4.0, 8.0\};}\nonumber
\end{eqnarray}
and then issue the command
\begin{eqnarray}
\text{In[9]:}&=\text{\em CheckAnalyNumer[u, n, $\mu$, $\nu$, uList, Iint] }\,,\nonumber
\end{eqnarray}
which returns a list of values containing the results of the symbolic algorithm and numerical calculations.
This is discussed in the next section.

\subsection{Validity of the results}\label{result-validity}
The extensive test file is one of the two files submitted to the electronic library. In this test which is carried out
over the practically important part of the problem domain, we did not encounter a single case where the
algorithm and its Mathematica implementation fail to hold. This extensive test was necessitated by both the present work and
the application mentioned in Ref. ~\cite{Gebremariam2}. In addition, as $n, \mu$ and $\nu$ increase, the instability of the naive numerical
computation becomes too prohibitive to use. This makes the symbolic integrator discussed in this paper and other numerically stable packages discussed in Refs. ~\cite{Singh, Deun} indispensable. Finally, we conclude this section with Table 1, that shows the comparison of the analytical and
numerical values for $I(u,-3,3,3)$ for a selected set of $u$ values. In both cases, the imaginary part is zero, hence it is not shown. As can be read from the table, the two values match exactly at least up to the reported number of decimal digits.
\begin{table}[ht]
\caption{$I(u, -3,3,3)$ } 
\centering 
\begin{tabular}{c c c} 
\hline\hline 
u & Analytical & Numerical \\ [0.5ex] 
\hline 
0.5   &  0.0000216   &  0.0000216  \\ 
1.0   &  0.0001523   &  0.0001523  \\
2.0   &  0.0005597   &  0.0005597 \\
4.0   &  0.0011940   &  0.0011940 \\
8.0   &  0.0017990   &  0.0017990 \\ [1ex] 
\hline 
\end{tabular}
\label{table:nonlin} 
\end{table}

\section{Comments and conclusions}
A symbolic algorithm for the analytical integration of a product of two spherical bessel functions with an additional
exponential and polynomial factor was discussed and implemented in a Mathematica package. The package complements existing numerical packages~\cite{Singh, Deun}.
It will especially be useful in situations where one has to calculate a huge number of these integrals analytically, as was the case in the application of the density matrix expansion ~\cite{Gebremariam2, Gebremariam3, Gebremariam4} to the HF energy obatined from chiral EFT two- and three-nucleon interactions at N$^2$LO. Possible directions of research
could be to expand the package to treat general bessel functions, to allow for the addition of a scaling parameter in the arguments of the bessel functions and to handle
more than two bessel functions as is done in the numerical approach of Ref. ~\cite{Deun}.
\section{Acknowledgements}
We thank R.~J.~Furnstahl for useful discussions. This work was supported in part by the National
Science Foundation under Grant No. PHY-0758125 and the UNEDF
SciDAC Collaboration (DOE Grant DE-FC02-07ER41457).

\section{Appendix}
\subsection{Analytic evaluation of $h_1(u)$ and $h_2(u)$}\label{appendix1}
Let us calculate the limits given in Eqs.\eqref{h1} and \eqref{h2}
\begin{eqnarray}
h_1 (u) &= &\lim_{y\to\infty}\biggl[ Ei\biggl(-i 2 y -\frac{y}{u}\biggr) +  Ei\biggl(i 2 y-\frac{y}{u}\biggr)
-2 \, Ei\biggl(-\frac{y}{u}\biggr)\biggr]\label{h1}\,,\\
h_2 (u)  &=& \lim_{y\to\infty}\biggl[ Ei\biggl(-i 2 y -\frac{y}{u} \biggr) -  Ei\biggl(i 2 y-\frac{y}{u} \biggr)\biggr]\label{h2} \,.
\end{eqnarray}
Starting with the definitions of the exponential integral function
given in Eq. \eqref{expinteg} and given its analyticity over the complex plane\footnote{It has a branch cut along $] -\infty, 0]$. }, performing a change of variable followed by a few simplification steps yield
\begin{eqnarray}
h_1(u) &=&\lim_{y\to\infty}\biggl[ e^{-y/u}\, \int^{i\, 2 y }_0 \frac{e^{-t}}{t + y/u} dt
\, + \, e^{-y/u}\, \int^{-i\, 2 y }_0 \frac{e^{-t}}{t + y/u} dt \biggr] \,,\\
h_2(u) &=&\lim_{y\to\infty}\biggl[ e^{-y/u}\, \int^{-i\, 2 y }_{i \, 2 y} \frac{e^{-t}}{t + y/u} dt \biggr]\,.
\end{eqnarray}
The next several steps involve change of variable of integration, applied uniformly to both $h_1(u)$
and $h_2(u)$ through rotation by $\pi/2$ and
$\pi$ consecutively. Hence
\begin{eqnarray}
h_1(u) &=&\lim_{y\to\infty}\biggl[ e^{-y/u}\, \int^{2 y }_0 \biggl(\, \frac{2 t \,\text{cos}\,t}{t^2 + (y/u)^2} \, + \,
\frac{2 y \,\text{sin}\,t}{u \bigl(t^2 + (y/u)^2\bigr)} \,\biggr) dt  \, \biggr] \,,\\
h_2(u) &=&\lim_{y\to\infty}\biggl[-i \,e^{-y/u}\, \int^{- 2 y }_{2 y} \frac{e^{i t }}{-i \,t + y/u} dt
\biggr]\,.
\end{eqnarray}
Hence, taking the limit of $h_1(u)$ directly and performing the remaining integration of $h_2(u)$ (the real and imaginary parts separately), followed by taking the
limit, we obtain
\begin{eqnarray}
h_1(u) &=& 0\,,\\
h_1(u) &=& - i \, 2 \pi\,.
\end{eqnarray}


\label{}



\end{document}